\documentclass{article}
\usepackage{spconf,amsmath,graphicx, mathtools, amsthm}
\usepackage{amsmath}
\usepackage{amssymb}
\usepackage{graphicx}
\usepackage{float}
\usepackage{booktabs}
\usepackage{array}
\usepackage{makecell}
\usepackage{multirow}
\usepackage{adjustbox}
\usepackage{subcaption}
\usepackage{algorithm}
\usepackage{algorithmic}
\usepackage{url}
\usepackage[colorlinks=true, linkcolor=blue, citecolor=blue, urlcolor=blue]{hyperref}

\title{Adaptive Voxelization for Transform coding of 3D Gaussian splatting data}
\name{Chenjunjie Wang$^*$, Shashank N. Sridhara$^*$, Eduardo Pavez$^*$, Antonio Ortega$^*$, Cheng Chang$^\dagger$\thanks{This work was funded in part by a gift from Meta. Source code available at: \url{github.com/STAC-USC/3DGS_Compression_Adaptive_Voxelization}}}
\address{$^*$University of Southern California, Los Angeles, CA\\
$^\dagger$Meta, Menlo Park, CA}
\long\def\comment#1{}

\newfont{\bbb}{msbm10 scaled 700}

\newfont{\bb}{msbm10 scaled 1000}

\newcommand{\qv}{{\bf q}}

\newcommand{\sv}{{\bf s}}

\newcommand{\Cm}{{\bf C}}

\newcommand{\Gm}{{\bf G}}

\newcommand{\Id}{{\bf I}}

\newcommand{\Nc}{{\cal N}}

\newcommand{\muv}{\hbox{\boldmath$\mu$}}

\newcommand{\Sigmam}{\hbox{\boldmath$\Sigma$}}

\begin{document}
\ninept
\maketitle
\begin{abstract}
We present a novel compression framework for 3D Gaussian splatting (3DGS) data that leverages transform coding tools originally developed for point clouds. 
Contrary to existing 3DGS compression methods, 
our  approach can produce compressed 3DGS models at multiple bitrates in a computationally efficient way. 
Point cloud voxelization is a discretization technique that point cloud codecs use to improve coding efficiency while enabling the use of fast transform coding algorithms. 
%
We propose an adaptive voxelization algorithm tailored to 3DGS data, to avoid the inefficiencies  introduced by uniform  voxelization used in point cloud codecs. We ensure the positions of larger volume Gaussians are represented at high resolution, as these significantly impact rendering quality. Meanwhile, a low-resolution representation is used for dense regions with smaller Gaussians, which have a relatively lower impact on rendering quality. This adaptive voxelization approach significantly reduces the number of Gaussians and the bitrate required to encode the 3DGS data. 
After voxelization, many Gaussians are moved or eliminated. Thus, we propose to fine-tune/recolor the remaining 3DGS attributes with an initialization that can reduce the amount of retraining required. 
Experimental results on pre-trained datasets show that our proposed compression framework outperforms existing methods.
\end{abstract}

%

\section{Introduction}
\label{sec:intro}
3D Gaussian Splatting (3DGS) is a state-of-the-art image-based 3D scene representation method \cite{kerbl2023_3dgs}. 
3DGS can produce high-quality models with a fraction of the training time while allowing faster rendering than its precursors---implicit radiance fields, such as NeRF's \cite{mildenhall2021nerf} and Instant-NGP \cite{M_ller_2022},  and Plenoxels \cite{Fridovich2022_plenoxels}. 3DGS provides an explicit representation of 3D scenes \cite{kerbl2023_3dgs}, consisting of a list of Gaussians to describe the scene geometry, each represented by (i) a mean vector (position), (ii) a covariance matrix, and (iii) appearance parameters to represent view-dependent colors,  i.e., opacity and spherical harmonic (SH) coefficients.
While the usage of 3DGS is expected to grow \cite{Fei2024_3dgssurvey}, the memory required to store millions of optimized Gaussian primitives makes compression crucial for efficient storage and transmission of 3DGS data in applications such as immersive communications and real-time streaming of 3D content \cite{bagdasarian2024_3dgssurvey}.

Existing works on 3DGS compression can be classified into \textit{generative} and \textit{post-training}  \cite{huang2024_ptc3dgs}. \textit{Generative} compression aims to reduce the memory footprint by producing lightweight 3DGS models through entropy-constrained training  and removal of redundant Gaussians  \cite{fan2024_lightgaussian, niedermayr2024_compressed3dgs, chen2024_hac3dgs, navaneet2023_comp3dgs}. 
Although \textit{generative} compression achieves significant reductions in memory while preserving rendering quality,  it requires retraining a different model for each bit-rate and rendering quality (rate-distortion points), which can be impractical for some applications where multiple rates need to be used. 

\textit{Post-training} compression methods \cite{huang2024_ptc3dgs, yang2024_GGSC, spz} take an existing pre-trained model and 
compress it without further retraining. 
Some of these methods \cite{huang2024_ptc3dgs, yang2024_GGSC} rely on point cloud compression (PCC) \cite{Graziosi_Nakagami_Kuma_Zaghetto_Suzuki_Tabatabai_2020, tmc13} because 3DGS data can be viewed as a 3D point cloud (3DPC), with Gaussian means as positions, and covariances, SH coefficients, and opacities as attributes. 
Post-training compression using PCC requires 
``voxelization'', where the Gaussian positions obtained from training are mapped to discrete voxels, analogous to pixels in images \cite{Graziosi_Nakagami_Kuma_Zaghetto_Suzuki_Tabatabai_2020}. After voxelization, the 3DGS points can be represented by an octree \cite{octree}, the data structure upon which all efficient PCC algorithms are built \cite{Graziosi_Nakagami_Kuma_Zaghetto_Suzuki_Tabatabai_2020,de2016compression}. 
Post-training codecs can produce 3DGS models at multiple bitrates and quality levels, similar to the established video/PC codecs. 
However, because relatively small distortions in 3DGS parameters (due to voxelization or transform coding) can lead to large distortions during rendering, \textit{post-training} methods require higher bitrates than \textit{generative} approaches to achieve similar quality levels.

\begin{figure}[t]
    \centering
    \begin{subfigure}[t]{0.49\linewidth}
        \centering
        \includegraphics[width=\linewidth]{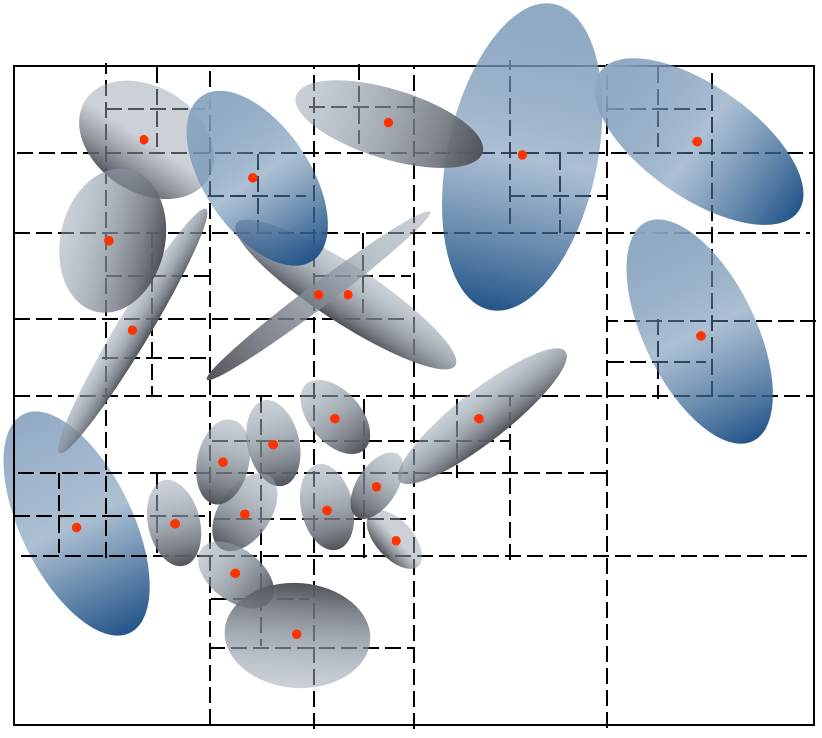}
        \caption{Uniform voxelization}
        \label{fig:uni_vox}
    \end{subfigure}
    \hfill
    \begin{subfigure}[t]{0.49\linewidth}
        \centering
        \includegraphics[width=\linewidth]{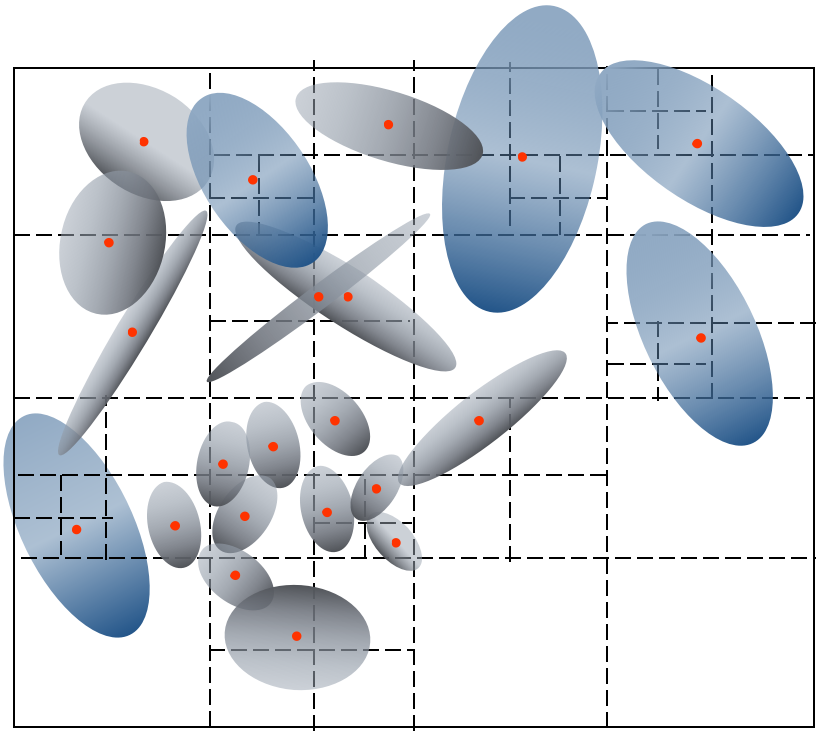}
        \caption{Adaptive voxelization}
        \label{fig:adap_vox}
    \end{subfigure}
    \caption{Voxelizations for  3DGS data with uniform resolution for all  positions (a), and with  resolution adaptive to the Gaussian volume and distribution (b).}
    \label{fig:vox_comparison}
\end{figure}

In this work, we propose a novel \textit{hybrid} compression framework that combines the strengths of \textit{generative} and \textit{post-training} compression, 
enabling us to achieve high rendering quality at bitrates comparable to \textit{generative} codecs while maintaining the computational efficiency and flexibility to produce compressed 3DGS models at multiple bitrates, as offered by \textit{post-training} codecs. 
\begin{figure*}[t]
    \centering
    \includegraphics[width=\textwidth, keepaspectratio]{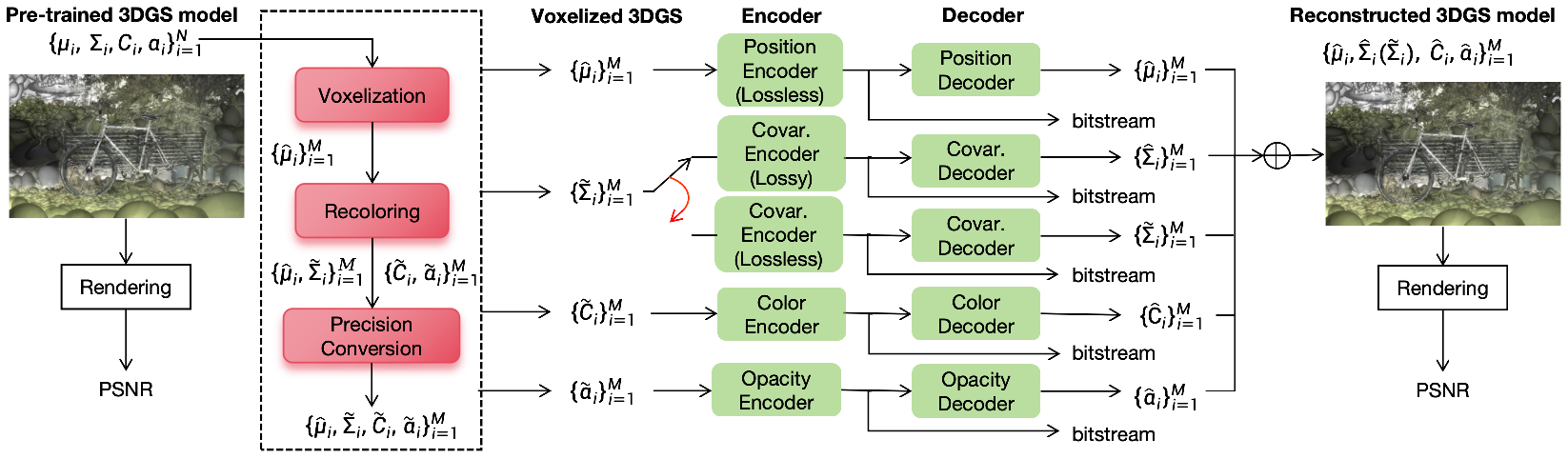}
    \caption{Proposed hybrid compression framework}
    \label{fig:pipeline}
\end{figure*}
Similar to \textit{post-training} methods, we leverage geometric point cloud compression (GPCC) techniques, which require the 3DGS positions to be voxelized. However, we introduce an additional stage of light re-training of attributes \textit{after voxelization and before compression}, which improves the rendering quality of the compressed model. 

While a 3DGS model can be viewed as a 3DPC with multiple types of attributes, there are several fundamental differences.  
First, 3DPCs from volumetric datasets have dense point distributions, while 3DGS point distributions can be irregular with dense and sparse regions. 
Second, in 3DPCs, each point contributes equally to the rendered images, while in 3DGS, larger volume Gaussians can influence larger regions in the rendered images. 
In uniform voxelization, such as that used in GPCC and in some 
post-training 3DGS compression approaches \cite{huang2024_ptc3dgs, xie2024_mesongs}, all regions of space are given the same position resolution, determined by the maximum depth of the octree.
Thus, uniform voxelization approaches can introduce large errors by producing large displacements in high-volume Gaussians while spending excessive bitrate in dense regions containing low-volume Gaussians. 
To address these challenges, we propose an \textit{adaptive voxelization for 3DGS data}, which selects higher resolution for regions containing large-volume Gaussians, or regions with higher voxelization errors,  and lower resolution for regions with dense Gaussians, as shown in \autoref{fig:adap_vox}.
While adaptive voxelization has been used to handle non-uniform 3DPC geometries \cite{zhang2020polarnet, sridhara2021_cylindrical}, our voxelization criteria are specific to 3DGS. 
After voxelization, Gaussians assigned to the same voxel can be merged, resulting in a significant reduction in the number of Gaussians, but requiring attribute (covariance, SH, and opacity) merging. 
A direct approach would be to retrain the reduced model from scratch, incurring significant retraining costs. Instead, we propose to initialize the attributes, followed by fine-tuning using the 3DGS learning algorithm from \cite{kerbl2023_3dgs}. 

In our proposed compression approach (see \autoref{fig:pipeline}), we first adaptively voxelize the 3DGS data. Next,  we apply a constrained, lightweight retraining of the 3DGS model attributes after voxelization. Then, the 3DGS data can be compressed using the MPEG GPCC codec \cite{tmc13, Graziosi_Nakagami_Kuma_Zaghetto_Suzuki_Tabatabai_2020,mpeg}. For covariance compression, we also use the vector quantization method from \cite{niedermayr2024_compressed3dgs}.
In what follows, \autoref{sec:compression_pipeline} describes
our hybrid compression framework, while \autoref{sec:adap_vox} presents our adaptive voxelization. Experiments
and conclusions are in \autoref{sec:experiment} and \autoref{sec:conclusion}.

\section{Compression pipeline}
\label{sec:compression_pipeline}
\subsection{3D Gaussian Splatting data}
3DGS are an explicit point-based 3D scene representation \cite{kerbl2023_3dgs}. The $i$th Gaussian   has a  mean or position $\muv_{i} \in \mathbb{R}^{3}$, a  covariance matrix $\Sigmam_{i} \in \mathbb{R}^{3 \times 3}$, $D$ spherical harmonic coefficients $\Cm_{i} \in \mathbb{R}^{D \times 3}$ to represent view-dependent color and opacities $\alpha_{i} \in [0, 1]$. Thus, a 3DGS model with $N$ Gaussians can be described as a list,
\begin{equation}
    \Gm = \{\muv_{i}, \Sigmam_{i}, \Cm_{i}, \alpha_{i} \}_{i = 1}^{N}.
\end{equation}
The 3DGS model $\Gm$ is learned from a set of training images $\{\tilde{\Id}_{i}\}_{i= 1}^{k}$, with camera view parameters $\{\tilde{\theta}_i\}_{i = 1}^{k}$, while the quality of a compressed 3DGS model $\hat{\Gm}$ is evaluated  using a different set of test images $\lbrace \Id_i \rbrace_{i=1}^m$, with corresponding camera views $\lbrace \theta_i \rbrace_{i=1}^m$. 
The losses used for training and testing are given by: 
\begin{align}\label{eq_test_distortion}
   \resizebox{.9\hsize}{!}{
   $\mathcal{L}_{\text{train}} =  \frac{1}{k}\sum_{i=1}^k \Vert R_{\tilde{\theta}_i}(\Gm) - \tilde{\Id}_i \Vert_F^2, \mathcal{L}_{\text{test}} = \frac{1}{m}\sum_{i=1}^m \Vert R_{{\theta}_i}(\hat{\Gm}) - {\Id}_i \Vert_F^2$,
   }
\end{align}
where $R_{\theta}$ is the rendering function on the view direction $\theta$ \cite{kerbl2023_3dgs}.
Note that changes in the 3DGS model affect the losses  $\mathcal{L}_{\text{train}}$ and $\mathcal{L}_{\text{test}}$ in \eqref{eq_test_distortion} through the rendering function. 

\subsection{Overview of proposed hybrid compression framework}
 Our proposed compression system is shown in \autoref{fig:pipeline}. 
 First, we voxelize the Gaussians' mean positions ($\muv_{i}$). Because many Gaussian means are quantized to the same voxel center, we remove repeated values and denote these new positions by $\{\hat{\muv}\}_{i = 1}^{M}$, where $M < N$. The remaining attributes are initialized at each voxel center $\{\Sigmam_i, \Cm_i, \alpha_i\}_{i = 1}^{N}$ and fine-tuned by minimizing $\mathcal{L}_{\text{train}}$ in \eqref{eq_test_distortion} until matching the quality of the input model $\Gm$,  while keeping the quantized mean positions fixed, resulting in a  voxelized 3DGS
     $\tilde{\Gm}=\{\hat{\muv}_{i}, \tilde{\Sigmam}_{i}, \tilde{\Cm}_{i}, \tilde{\alpha}_{i} \}_{i = 1}^{M}.$
 Further details about voxelization and fine-tuning are provided in \autoref{sec:adap_vox}. 
$\tilde{\Gm}$ is a voxelized point cloud with multiple attributes, which can be compressed using GPCC.  

\subsection{Position and covariance compression}

The voxelized positions $\lbrace \hat{{\muv}_i} \rbrace_{i=1}^M$ are losslessly compressed using GPCC, which uses octree coding followed by an octree-designed entropy coder \cite{tmc13}. Further details about uniform and adaptive voxelization are provided in \autoref{sec:adap_vox}.

For covariance compression  $\lbrace\tilde{\Sigmam}_i\rbrace_{i=1}^M$, we employ two strategies based on the desired balance between compression ratio and geometric fidelity: \emph{lossy compression} using vector quantization  (VQ)~\cite{niedermayr2024_compressed3dgs} and transform-based {lossless compression} with GPCC ~\cite{tmc13, Graziosi_Nakagami_Kuma_Zaghetto_Suzuki_Tabatabai_2020}. 
In both cases, the covariance matrix $\tilde{\Sigmam}_i$ is represented by a quaternion vector $\qv_{i} \in \mathbb{R}^{4}$ and a  scale vector $\sv_{i} \in \mathbb{R}^{3}$. The quaternion vector represents the orientation of the Gaussian, while the scale vector represents its size.  
The \emph{lossy covariance compression} method from \cite{niedermayr2024_compressed3dgs} designs a codebook based on the relative impact of each Gaussian on the rendered images. 
For \emph{lossless covariance compression}, we configure GPCC to use the predictive lifting transform to compress each dimension of the quaternion and scale vectors independently, as if they were point cloud intensities. 
The VQ-based method offers a higher compression ratio at the cost of complexity and distortion, while the transform-based method guarantees lossless reconstruction but with a higher bitrate.


\subsection{SH coefficients and opacity compression}
SH coefficients are used to represent view-dependent color, with the low-frequency component ($\Cm_{i}^{(\mathrm{dc})} \in \mathbb{R}^{1 \times 3}$) representing the base color and the high-frequency components ($\Cm_{i}^{(\mathrm{ac})} \in \mathbb{R}^{15 \times 3}$) capturing view-dependent variations. Since the rendered RGB color is a linear combination of SH coefficients and SH basis functions, 
they can be similarly converted to the YUV domain to de-correlate the color channels before transform coding \cite{shashank2024_mpeg3dgs}.
%
After transforming the SH coefficients to the YUV domain, we compress them by  GPCC configured to run the Region Adaptive Hierarchical Transform (RAHT) for each frequency as if they were YUV colors. RAHT is a multi-resolution transform designed to remove spatial correlation in point cloud color attributes \cite{de2016compression}. For each frequency, the   transformed YUV coefficients  are uniformly quantized and jointly entropy coded using the run-length Golomb-Rice (RLGR) algorithm from GPCC, in order to exploit redundancies across color channels. Opacity is compressed similarly, but using single-channel intra RAHT instead  \cite{tmc13, Graziosi_Nakagami_Kuma_Zaghetto_Suzuki_Tabatabai_2020}. 
We generate rate-distortion (R-D) points by using quantization parameters ($Q_{dc}, Q_{ac},Q_{op}$) for transform coding of $\Cm_{i}^{(\mathrm{dc})}$, $\Cm_{i}^{(\mathrm{ac})}$, and  $\alpha_{i}$, respectively. Since $\Cm_{i}^{(\mathrm{dc})}$ and $\alpha_{i}$ are more critical for rendering, and  $\Cm_{i}^{(\mathrm{ac})}$ contains significant redundancy, we set $Q_{dc}$ and $Q_{op}$ to be relatively smaller than $Q_{ac}$ in GPCC.

\begin{figure}[t]
    \centering


    \begin{minipage}{0.33\linewidth}
        \centering
        \adjustbox{width=\linewidth, height=1.7cm, trim=15 15 15 15, clip}{\includegraphics{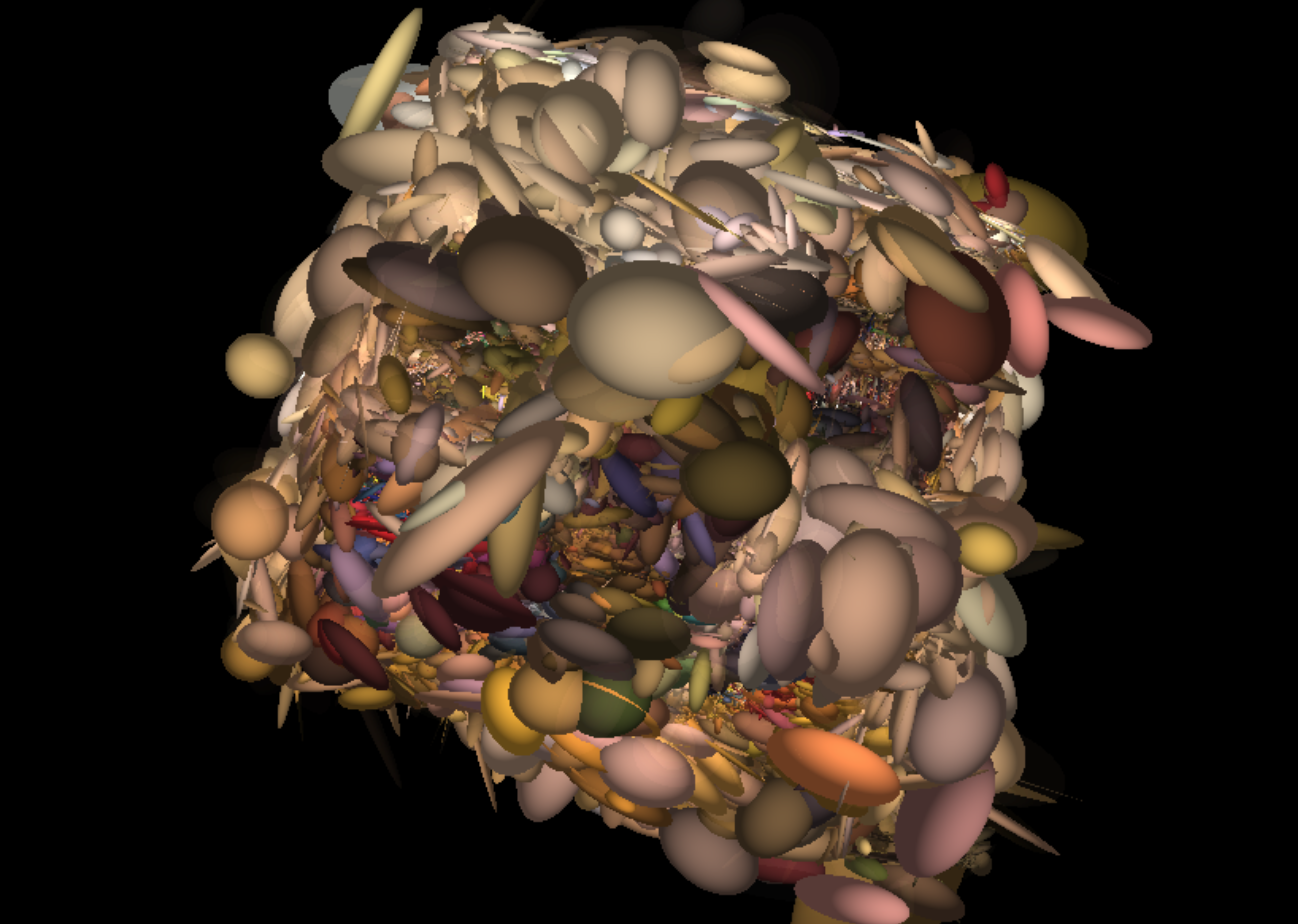}}
        \subcaption{Playroom (Bounded)}
        \label{fig:bounded}
    \end{minipage}
    \hspace{0.125cm}
    \begin{minipage}{0.38\linewidth}
        \centering
        \adjustbox{width=\linewidth, height=1.7cm, trim=10 10 10 10, clip}{\includegraphics{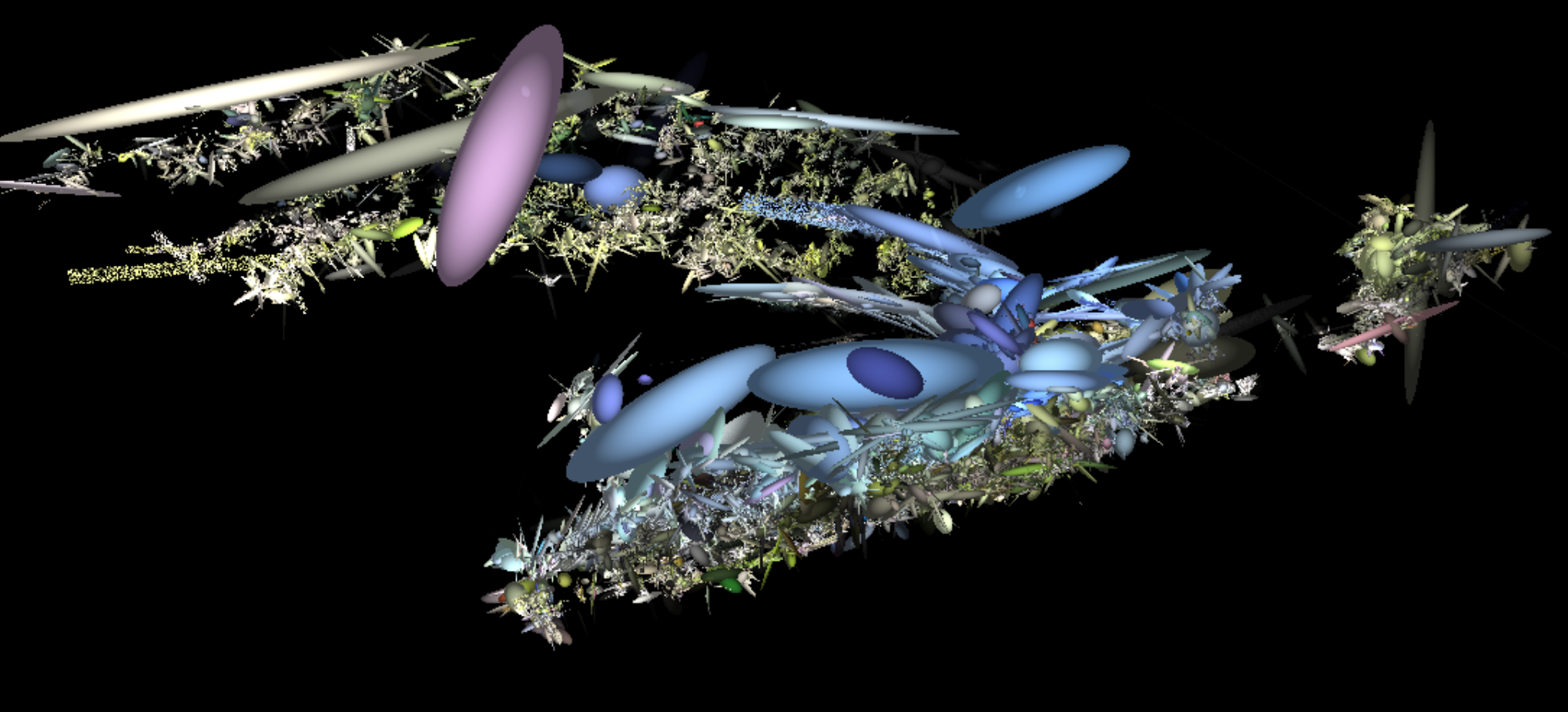}}
        \subcaption{Train (Unbounded)}
        \label{fig:unbounded}
    \end{minipage}
    \caption{Distribution of 3D Gaussians in bounded and unbounded scenes from pre-trained models.}
    \label{fig:scene_combined}
\end{figure}

\section{Voxelization}
\label{sec:adap_vox}

\subsection{Uniform voxelization}
\begin{figure}[t] 
    \centering
    \includegraphics[width=0.96\linewidth]{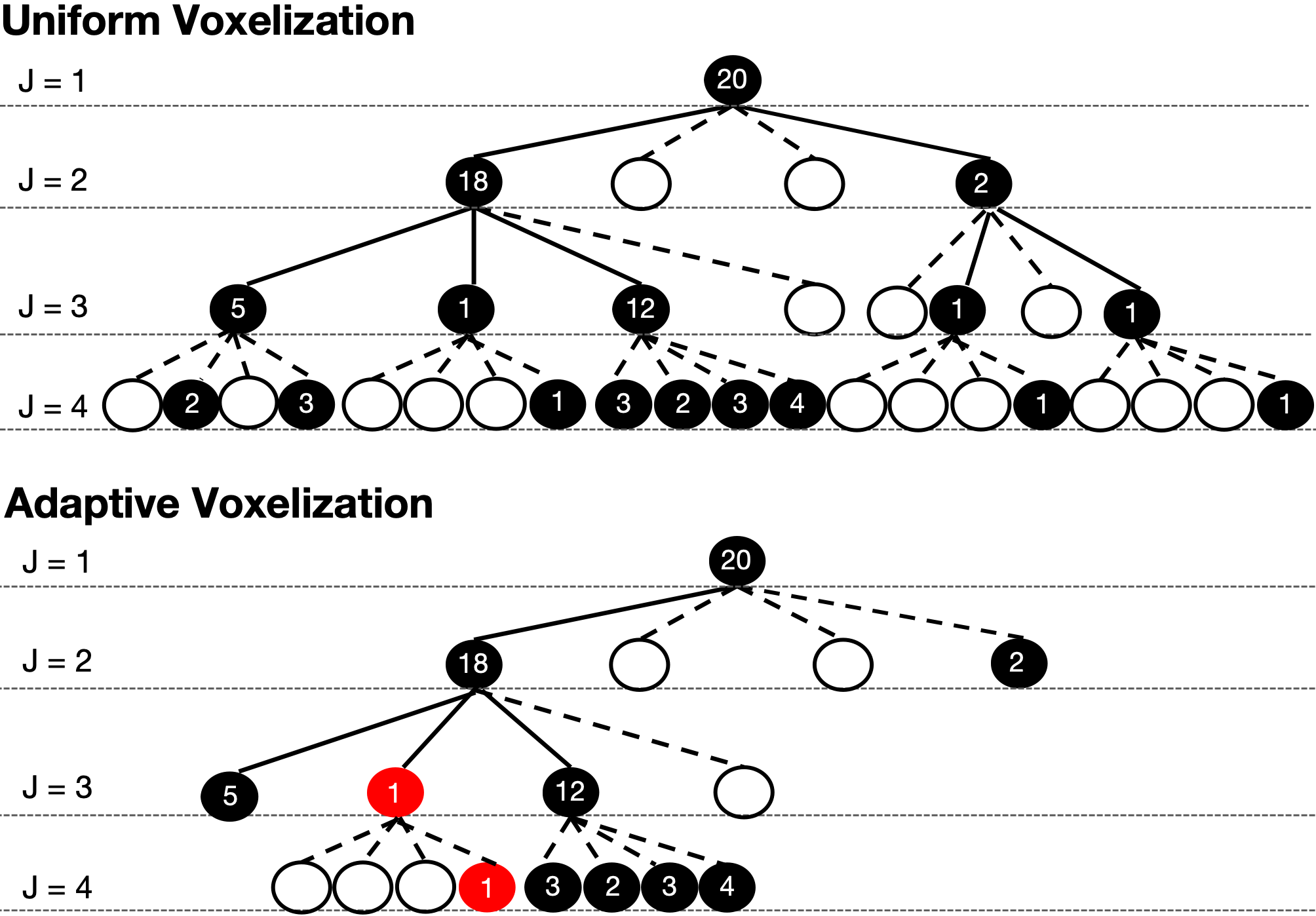}
    \caption{Partitioning tree structure voxelization. 
    \raisebox{-1.5pt}{\includegraphics[height=0.9\baselineskip]{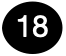}}
    indicates a voxel with 18 Gaussian centers, while 
    \raisebox{-1.5pt}{\includegraphics[height=0.9\baselineskip]{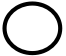}}
    represents an empty sub-voxel. 
    Uniform voxelization maintains a fixed resolution $J_{uni}=4$ for all occupied voxels, while adaptive voxelization adjusts resolution dynamically based on a threshold $\tau=5$ with $J_{low}=1$ and $J_{high}=4$. 
    \raisebox{-1.5pt}{\includegraphics[height=0.9\baselineskip]{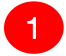}}
    marks voxels with large Gaussians.}
    \label{fig:Tree} 
\end{figure}

In uniform voxelization, the 3D space is partitioned into volume elements called \emph{voxels} (cubes). The positions of all points assigned to a voxel are quantized to the voxel center, and their corresponding attributes are averaged \cite{dynamic_polygon_cloud}. 
If the bounding box of the 3D points has a volume of $W \times W \times W$, the first level of the octree divides the volume into $2^3$ cubes, each with a volume $W/2 \times W/2 \times W/2$. At each subsequent level, only the occupied voxels are divided. 
After $J$ levels of partitioning, the resulting cubes have a volume of $W/2^J \times W/2^J \times W/2^J$, which represents the resolution of the voxelized 3D points. \autoref{fig:uni_vox} shows voxelization in 2D space, while \autoref{fig:Tree} shows the octree representation of voxelized 3D points to a specified depth. 
Uniform voxelization is efficient if points (Gaussian means) are uniformly distributed, but becomes inefficient in regions with non-uniform point density, such as those observed in unbounded 3DGS scenes (\autoref{fig:unbounded}). 
Additionally, it only considers the position but not the distribution or the size of the Gaussians in a voxel, thus ignoring the relative importance of each Gaussian for rendering. \autoref{fig:bounded} and \autoref{fig:unbounded} depict the variability in density and volume of the 3DGS data distribution.


\subsection{Adaptive Voxelization}
Adaptive voxelization is achieved by defining criteria to determine whether an occupied voxel should be further partitioned.

\subsubsection{Problem formulation}
Consider a voxel with $k$ Gaussians $\{G_i\}_{i=1}^{k}$, each parametrized by its mean and covariance, $G_{i} \sim \Nc(\muv_i, \Sigmam_i)$. 
During voxelization, these $k$ Gaussians are replaced with a single Gaussian, $G^{*} \sim \Nc(\muv^{*}, \Sigmam^{*})$, whose corresponding SH coefficients, opacities are later fine-tuned to minimize $\mathcal{L}_{\text{train}}$ in \eqref{eq_test_distortion}. Our criterion for determining whether to further partition a voxel is based on the ``geodesic distance'' between $G^{*}$ and $\{G_i\}_{i=1}^{k}$, which  
can be quantified using the \emph{2-Wasserstein distance}, $W_{2}^{2}(G_i, G_j)$ \cite{delon2020_wasserstein}:
\begin{equation}
\label{eqn:wass_dist}
\resizebox{.9\hsize}{!}{$ W_{2}^{2}(G_i, G_j) = \lVert \muv_i - \muv_j \rVert_{2}^{2} + \mathrm{Tr}(\Sigmam_i + \Sigmam_j - 2(\Sigmam_i^{1/2} \Sigmam_j \Sigmam_i^{1/2})^{1/2})$},
\end{equation}
where the first term is the distance between Gaussian means and the second term represents the similarity of the covariance matrices. We define the distance between $G^{*}$ and the original Gaussians $\{G_i\}_{i=1}^{k}$: 
\begin{equation}
    \label{eqn:w2_criterion}
    d = \sum_{i=1}^{k} W_{2}^{2}(G_i, G^*).
\end{equation}
The Gaussian $G^{*}$ with parameters $\muv^*$ and $\Sigmam^*$ that minimizes \eqref{eqn:w2_criterion} is called the $W_2$-Barycenter of  $\{G_i\}_{i=1}^{k}$ and is computed using \cite{delon2020_wasserstein}:  
\begin{align}
\resizebox{.9\hsize}{!}{$\muv^* = \frac{1}{k}\sum_{i = 1}^{k} \muv_i, \Sigmam^*(l+1) = \sum_{i = 1}^{k} ((\Sigmam^*(l))^{1/2}\Sigmam_{i} (\Sigmam (l))^{1/2})^{1/2}$}.
\end{align}
Solving for $\Sigmam^{*}$ requires a computationally expensive iterative algorithm \cite{alvarez2016_fixedsigma}. In what follows, we propose simpler criteria that do not require explicitly computing $\Sigmam^*$ and can reduce the runtime by approximately a factor of $15$.

\subsubsection{Proposed criteria for adaptive voxelization}
We consider three criteria to decide whether an occupied voxel should be further partitioned at each level: 1) \textit{volume} of the Gaussians, 2) \textit{number of Gaussians} in the voxel, and 3) \textit{distance of the Gaussians to voxel center}. These criteria are selected to minimize \eqref{eqn:w2_criterion} without explicitly computing $\Sigmam^*$.

\begin{figure}[t] 
    \centering
    \includegraphics[width=\linewidth]{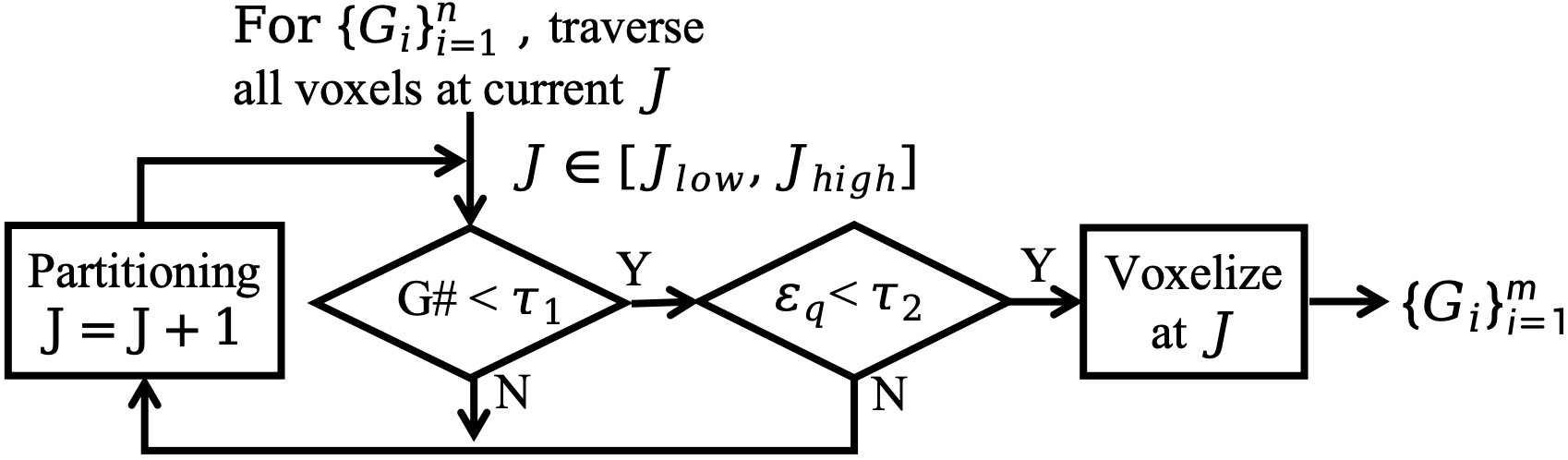}
    \caption{Large Gaussians are represented at the highest resolution. For the remaining Gaussians, at a specified depth, each voxel is further partitioned if: (1) if the number of Gaussians exceeds the threshold ($\tau_1$); or (2) if the distance of the original Gaussian means to the voxel center, is larger than half the voxel size ($\tau_2 = W/2^{J+1}$). }
    \label{fig:adap_flow} 
\end{figure}

First, we compute the volume of each Gaussian in the 3DGS data:  $V_i = \sv_i^{(x)} \times {\sv_i^{(y)}} \times {\sv_i^{(z)}}$. The largest $v\%$ Gaussians, based on volume, are represented at the highest possible resolution $J_{\mathrm{high}}$, as the 2-Wasserstein distance between a large volume Gaussian and its representative $G^*$ increases significantly due to the trace term in \eqref{eqn:wass_dist}. The remaining Gaussians are voxelized uniformly up to a depth of $J_{\mathrm{low}}$. Beyond this depth, further partitioning is determined based on the number of Gaussians within a voxel and their distance from the voxel center. If a voxel at a given depth contains a high number of Gaussians that are non-uniformly distributed, the metric in \eqref{eqn:w2_criterion} will increase due to the first term of the 2-Wasserstein distance, indicating the need for further partitioning. To this end, we define thresholds for the number of Gaussians ($\tau_{1}$) and distance to center ($\tau_{2}$) to check if the given voxel satisfies the criteria. Our proposed criteria for adaptive voxelization after the volume criterion is summarized in \autoref{fig:adap_flow}. 
%
Given  $J_{\mathrm{high}}$, an octree obtained from  \emph{adaptive voxelization} is a pruned version of the original uniformly voxelized octree with the same  $J_{\mathrm{high}}$.
Therefore, adaptive voxelization
leads to smaller octrees (see \autoref{fig:Tree}). 

\subsection{Fine-tuning/recoloring}
\label{subsec:finetuning}
After voxelizing the Gaussian means, the remaining attributes $\Sigmam_i, \Cm_i, \alpha_i$ are fine-tuned through a constrained retraining process that fixes the Gaussian means. Initializing the attributes to zero would take more iterations, and it often results in sub-optimal rendering quality, since Gaussian splitting and merging are disabled while retraining. Instead, we propose to initialize the attributes before fine-tuning. 
Similar to point cloud color attributes, we initialize the DC colors $\Cm_{i}^{(\mathrm{dc})}$, high-frequency SH coefficients $\Cm_{i}^{(\mathrm{ac})}$ and opacities $\alpha_i$ of the voxelized Gaussian means by averaging the attributes within each voxel. Because the $W_2$-Barycenter covariance $\Sigmam^*$ \cite{alvarez2016_fixedsigma} is dominated by larger Gaussians,  we initialize the covariance of the voxelized Gaussian mean by the largest volume Gaussian in the voxel.

\section{Experiments}
\label{sec:experiment} 
We evaluate the proposed compression pipeline on two scenes: a bounded indoor scene, \emph{Playroom}, from the Deep blending dataset \cite{DeepBlending2018}, and an unbounded outdoor scene, \emph{Truck}, from the Tanks and Temples dataset \cite{Knapitsch2017_tankstemples}. 
We first present baseline results using the proposed codec with uniform voxelization and fine-tuning without initialization. 
Next, we analyze the impact of initialization for fine-tuning and adaptive voxelization on compression. 
Finally, we compare our proposed codec with adaptive voxelization against current state-of-the-art 3DGS coding schemes. In all the experiments, we report the number of bits required to encode the 3DGS model and the PSNR of the rendered decoded 3DGS with respect to the test-set images following $\mathcal{L}_{\text{test}}$ in \eqref{eq_test_distortion}. Details about covariance compression and choices of quantization parameters for SH coefficients and opacities are described in  \autoref{sec:compression_pipeline}. 

\subsection{Evaluation of proposed codec}
\subsubsection{Baseline result}
\begin{figure}[t]
    \centering
    \begin{subfigure}[t]{0.49\linewidth}
        \centering
        \includegraphics[width=\linewidth]{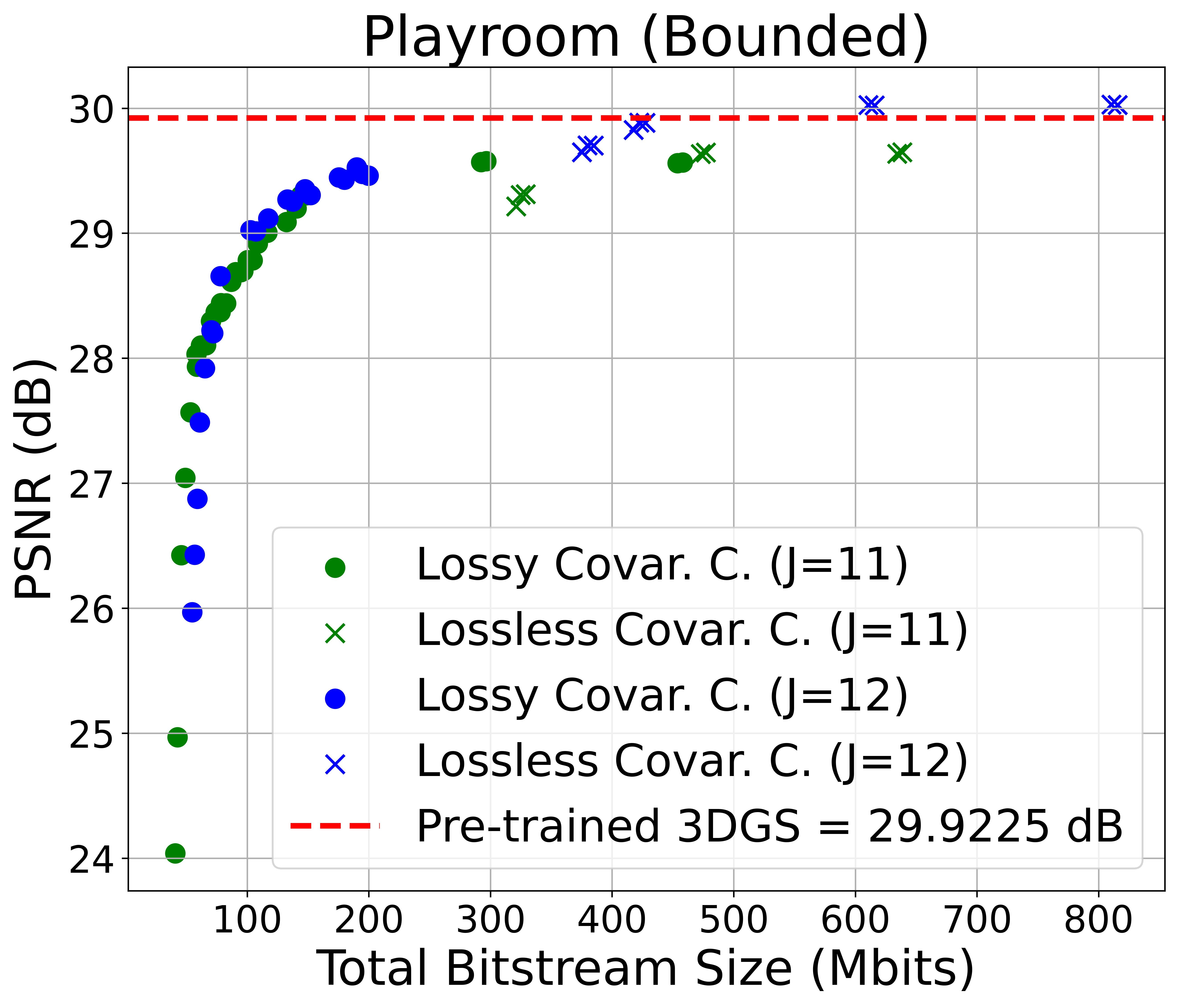}
        \caption{Playroom (Deep blending \cite{DeepBlending2018})}
        \label{fig:codec_playroom}
    \end{subfigure}
    \hfill
    \begin{subfigure}[t]{0.49\linewidth}
        \centering
        \includegraphics[width=\linewidth]{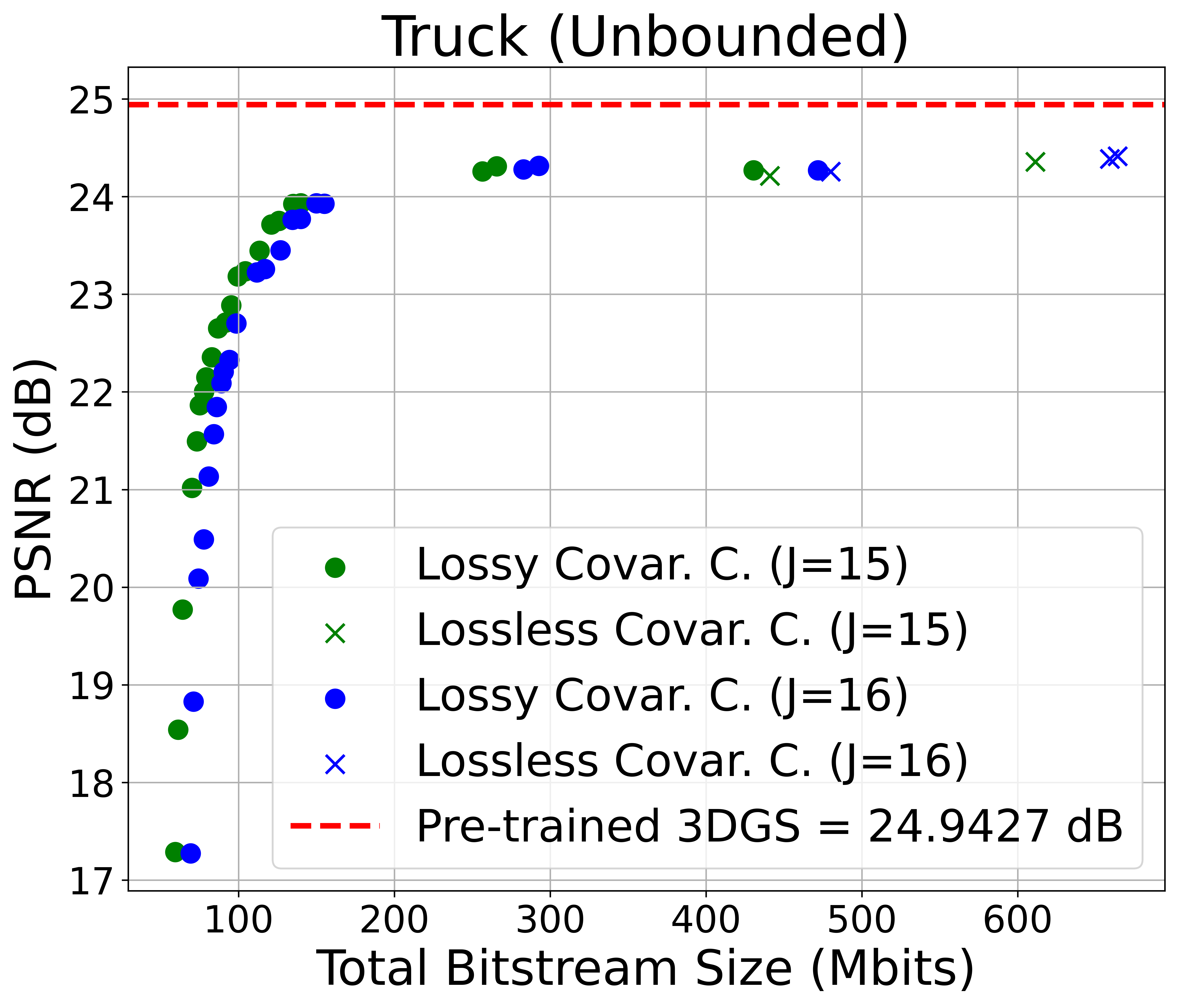}
        \caption{Truck (Tanks and temples \cite{Knapitsch2017_tankstemples})}
        \label{fig:codec_truck}
    \end{subfigure}
    \caption{Rate-distortion plots of proposed codec with uniform voxelization and fine-tuning without initialization on bounded (\autoref{fig:codec_playroom}) and unbounded (\autoref{fig:codec_truck}) scenes.}
    \label{fig:Baseline}
\end{figure}

We evaluate our proposed codec using uniform voxelization of Gaussian means ($\muv_{i}$) with different voxelization depths $J_{\mathrm{uni}}$, while fine-tuning the remaining attributes ($\Sigmam_{i}, \Cm_{i}, \alpha_{i}$) without initialization. We observe that increasing the voxelization depth $J_{\mathrm{uni}}$ helps retain the ground truth PSNR of the pre-trained model at higher bitrates for bounded scenes (\autoref{fig:codec_playroom}). However, at higher bitrates, the ground-truth PSNR cannot be fully recovered for the unbounded scene (\autoref{fig:codec_truck}) due to the inefficiency of uniform voxelization in regions where Gaussians are unevenly distributed.

\subsubsection{Effect of proposed initialization for fine-tuning}
\begin{figure}[t]
    \centering
    \begin{subfigure}[t]{0.49\linewidth}
        \centering
        \includegraphics[width=\linewidth]{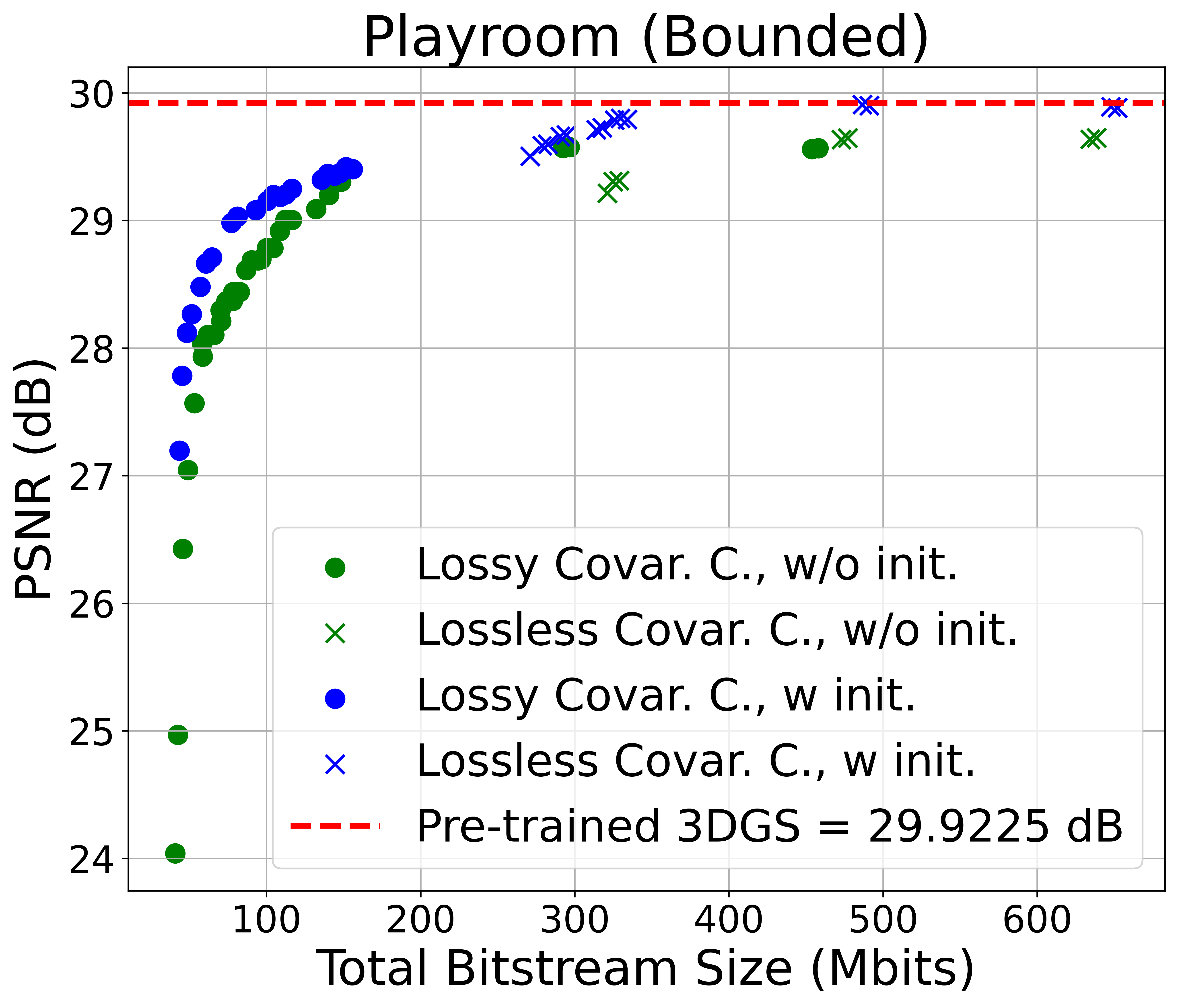}
        \caption{Playroom ($J_{\mathrm{uni}}=11$)}
        \label{fig:Retrain_playroom}
    \end{subfigure}
    \hfill
    \begin{subfigure}[t]{0.49\linewidth}
        \centering
        \includegraphics[width=\linewidth]{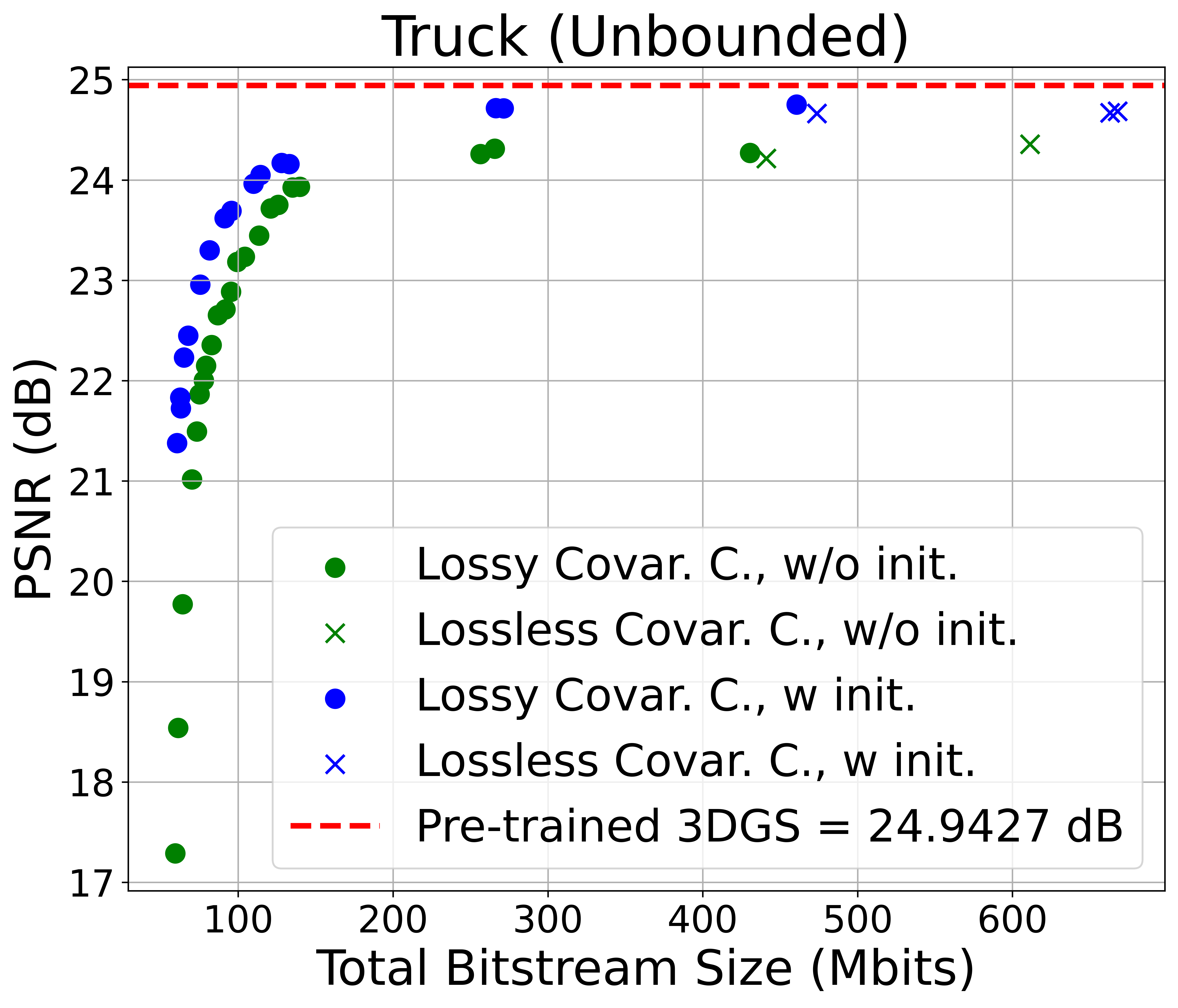}
        \caption{Truck ($J_{\mathrm{uni}}=15$)}
        \label{fig:Retrain_truck}
    \end{subfigure}
    \caption{Rate-distortion plots of proposed codec using uniform voxelization (fixed depth $J_{\mathrm{uni}}$) with and without proposed initialization for fine-tuning.}
    \label{fig:Retraining}
\end{figure}

We compare the impact of fine-tuning/recoloring the covariances ($\Sigmam_{i}$), SH coefficients ($\Cm_{i}$) and opacities ($\alpha_{i}$) after voxelization using our proposed initialization. $30$K iterations of fine-tuning without initialization using \cite{kerbl2023_3dgs}  fail to recover the ground-truth PSNR at higher rates. In contrast, our proposed initialization of $\Sigmam_{i}, \Cm_{i}, \alpha_{i}$ (see \autoref{subsec:finetuning}) requires only around $8$K iterations of fine-tuning to approach ground-truth PSNR for both bounded and unbounded scenes (\autoref{fig:Retrain_playroom} and \autoref{fig:Retrain_truck}).

\subsubsection{Effect of proposed adaptive voxelization}
\begin{figure}[t]
    \centering
    \begin{subfigure}[t]{0.49\linewidth}
        \centering
        \includegraphics[width=\linewidth]{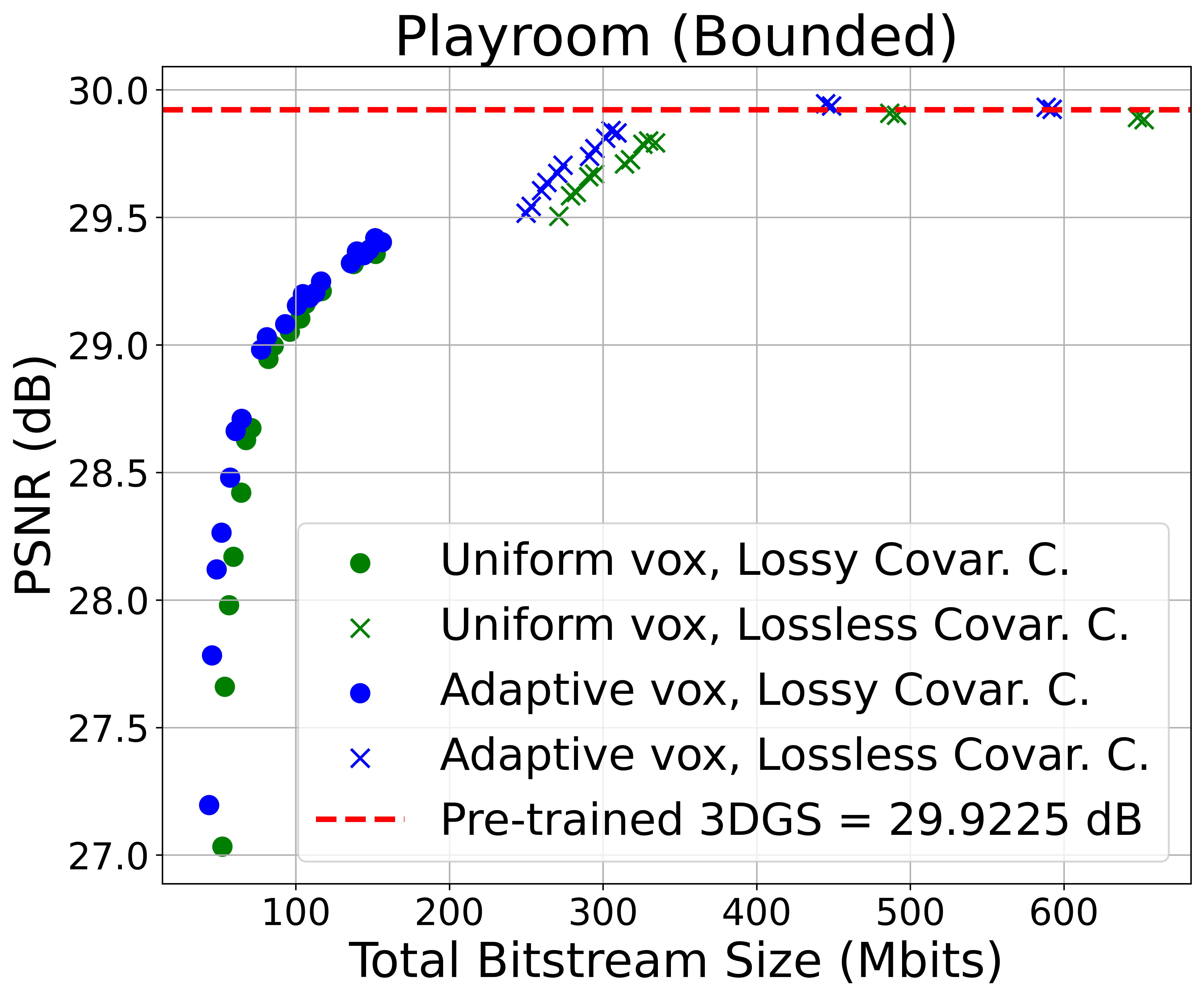}\subcaption{Adaptive:$J_{\mathrm{low}}=8$,$J_{\mathrm{high}}=16$}
        \label{fig:Adapt_playroom}
    \end{subfigure}
    \hfill
    \begin{subfigure}[t]{0.49\linewidth}
        \centering
        \includegraphics[width=\linewidth]{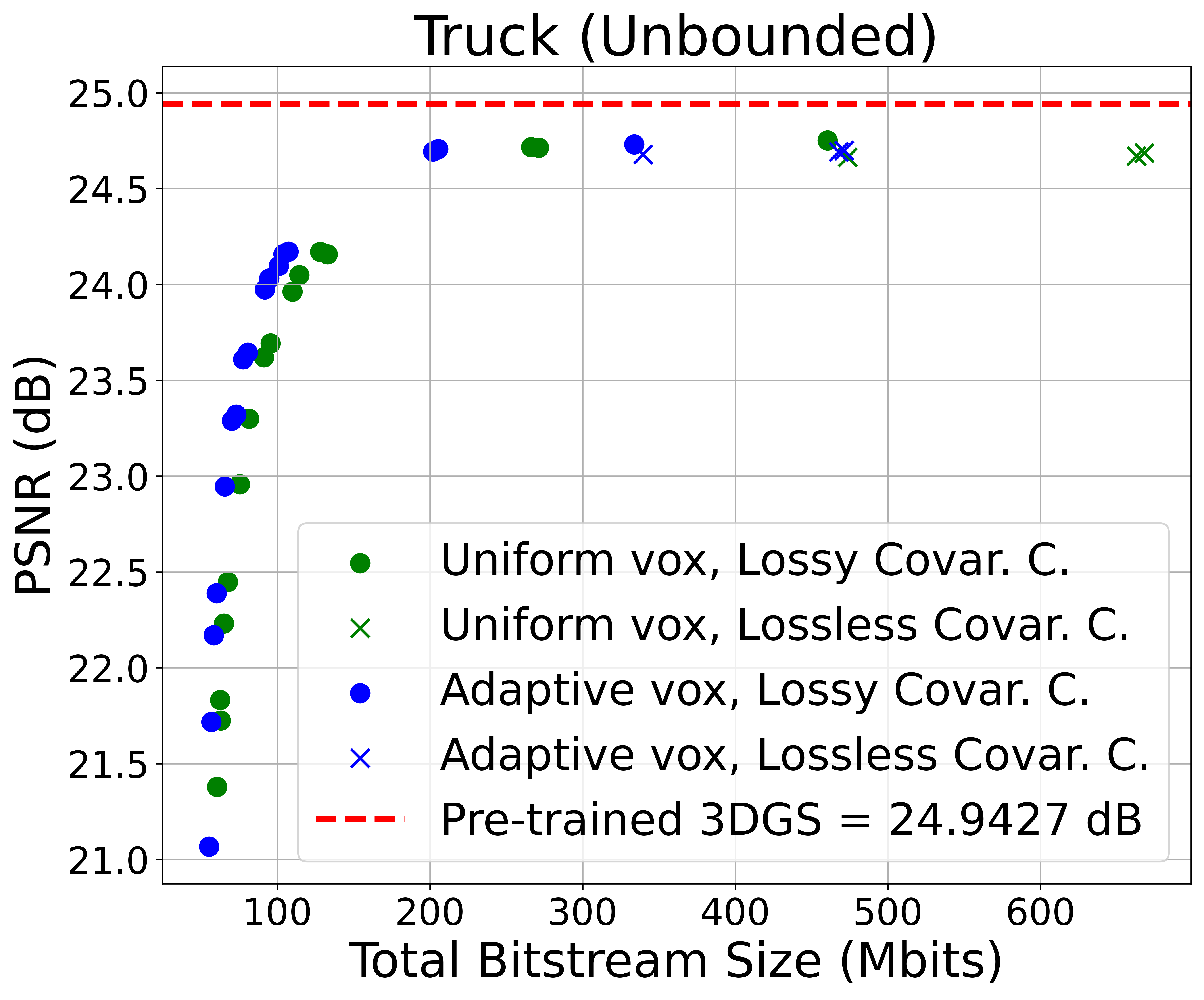}
        \subcaption{Adaptive:$J_{\mathrm{low}}=8$,$J_{\mathrm{high}}=18$}
        \label{fig:Adapt_truck}
    \end{subfigure}
    \caption{Comparison of rate-distortion plots of proposed codec using uniform voxelization (fixed depth $J_{\mathrm{uni}}$) and adaptive voxelization (variable depth from $J_{\mathrm{low}}$ to $J_{\mathrm{high}}$). Uniform voxelization depths for playroom and truck to $J_{\mathrm{uni}}=11$ and $J_{\mathrm{uni}}=15$, respectively.}
    \label{fig:Adaptive_combined}
\end{figure}

We compare uniform voxelization and our proposed adaptive voxelization in terms of compression performance. After voxelization, we apply the proposed initialization to fine-tune the remaining attributes and encode them. For uniform voxelization, we use a fixed depth $J_{\mathrm{uni}}$, while adaptive voxelization employs variable depths ranging from $J_{\mathrm{low}}$ to $J_{\mathrm{high}}$. As a result, the number of Gaussians to encode differs between uniform and adaptive voxelization. The compression results, shown in \autoref{fig:Adapt_playroom} and \autoref{fig:Adapt_truck}, demonstrate that adaptive voxelization improves compression performance for both bounded and unbounded scenes. This improvement can be attributed to the significant reduction in the number of Gaussians after adaptive voxelization compared to uniform voxelization. 
In \autoref{tab:bd_metric}, we report bitrate savings and average PSNR gain using the Bjontegaard metric \cite{bjontegaard2001}. The adaptive voxelization results in bitrate savings of up to $13\%$ for both bounded and unbounded scenes. 

\begin{table}[h]
\centering
\caption{Compression gains from adaptive voxelization compared to uniform voxelization}
\label{tab:bd_metric}
\begin{tabular}{|c|c|c|}
\hline
\textbf{scene} & \textbf{avg. PSNR gain} & \textbf{bitrate savings} \\ \hline
Playroom          & 0.08                    & 13.42\%                  \\ \hline
Truck             & 0.19                    & 12.64\%                  \\ \hline
\end{tabular}
\end{table}

\subsection{Comparison with state-of-the-art compression methods}
\begin{figure}[htbp]
    \centering
    \begin{subfigure}[t]{0.49\linewidth}
        \centering
        \includegraphics[width=\linewidth]{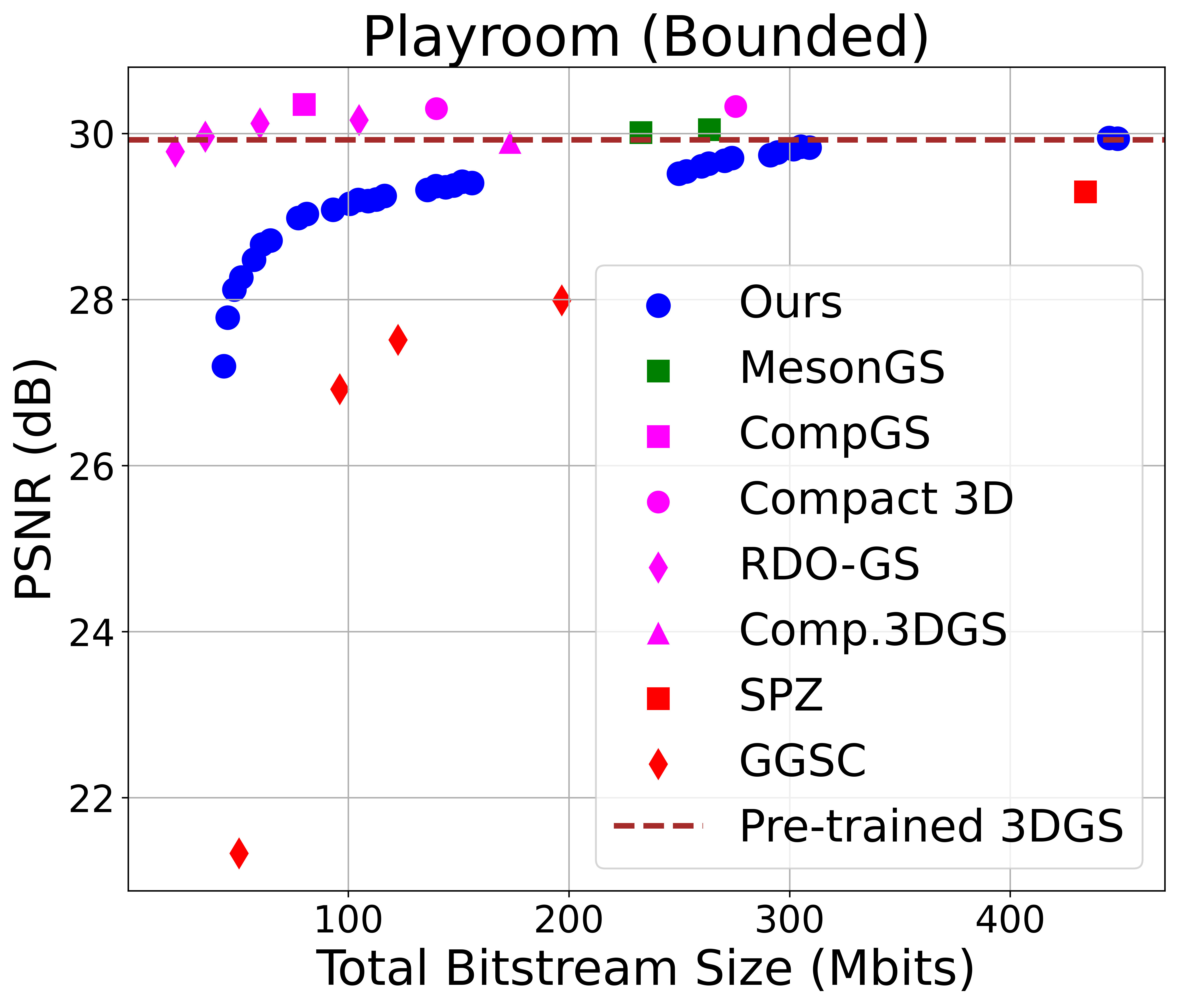}
        \caption{Retrain 3DGS}
        \label{fig:comparison_playroom}
    \end{subfigure}
    \hfill
    \begin{subfigure}[t]{0.49\linewidth}
        \centering
        \includegraphics[width=\linewidth]{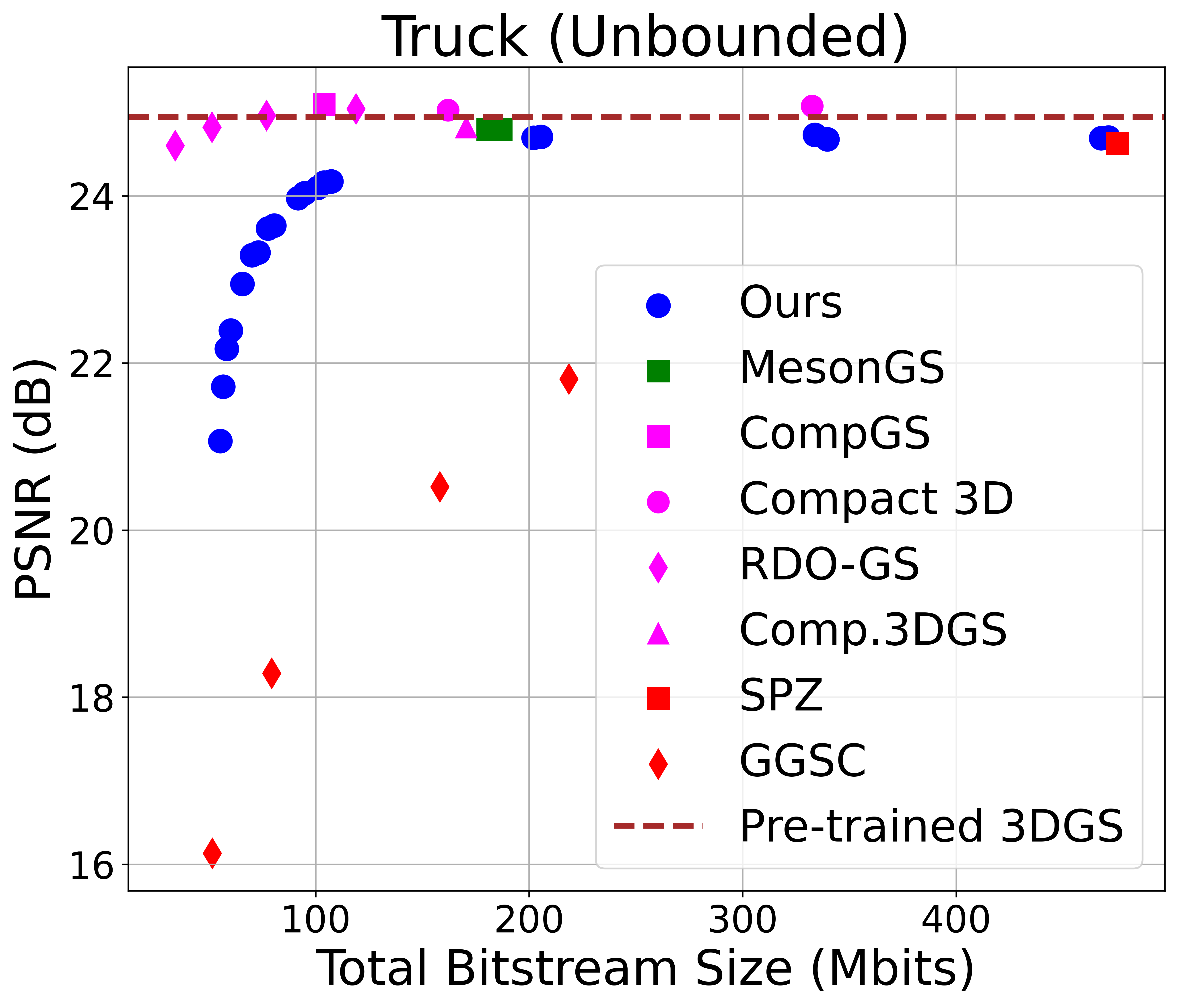}
        \caption{Retrain 3DGS}
        \label{fig:comparison_truck}
    \end{subfigure}
    \caption{Comparison of proposed codec using adaptive voxelization with state-of-the-art compression 3DGS methods. }
    \label{fig:comparison_sota}
\end{figure}

We conducted a comprehensive comparison of our proposed compression pipeline with adaptive voxelization against the state-of-the-art \emph{generative}, \emph{post-training} and \emph{hybrid} 3DGS compression methods (see  \autoref{tab:summary_sota}). 
From \autoref{fig:comparison_sota},  
our method outperforms post-training compression techniques: GGSC \cite{yang2024_GGSC} and SPZ \cite{spz}, by large margins. While generative compression methods such as CompGS \cite{navaneet2023_comp3dgs}, Compact 3D \cite{lee2024compact3dgaussianrepresentation}, RDO-GS \cite{wang2024_rdogs} and Compressed 3DGS \cite{niedermayr2024_compressed3dgs} achieve superior compression performance, they lack the ability to generate a compressed model at a specific bitrate, making rate control infeasible. 
In addition, they require retraining a completely different model for each R-D point. These limitations are critical in practical applications such as 3D data streaming. 
In contrast, our proposed hybrid codec enables the generation of multiple R-D points within a single encoding process. Our proposed method achieves performance comparable to the hybrid codec MesonGS \cite{xie2024_mesongs}, which involves fine-tuning after post-training compression. However, MesonGS requires fine-tuning at the decoder, which is impractical as it necessitates storing or sending the training images to the decoder. As a result, it can generate only a single R-D point per encoding-decoding process.

\begin{table}[ht]
\centering
\caption{Summary of current state-of-the-art 3DGS compression. GC (Generative Compression), PTC (Post-Training Compression), HC (Hybrid Compression), FT (fine-tuning), VQ (Vector Quantization), and GFT (Graph-Fourier Transform)}
\begin{adjustbox}{width=0.45\textwidth,center}
\label{tab:summary_sota}
\begin{tabular}{|l|c|c|c|c|c|}
\hline
\textbf{Method} & \textbf{Type} & \textbf{Enc} & \textbf{Dec} & \textbf{\begin{tabular}[c]{@{}c@{}}retrain\\ itr.\end{tabular}} & \textbf{\begin{tabular}[c]{@{}c@{}}r-d pts/\\ retrain\end{tabular}} \\ \hline
CGS \cite{navaneet2023_comp3dgs}           & GC            & VQ               & NA               & 30K                                                            & 1                                                                         \\ \hline
C3D \cite{lee2024compact3dgaussianrepresentation}     & GC            & VQ               & NA               & 30K                                                            & 1                                                                         \\ \hline
RDOGS \cite{wang2024_rdogs}       & GC            & VQ               & NA               & 30K                                                            & 1                                                                         \\ \hline
C-3DGS \cite{niedermayr2024_compressed3dgs}     & GC            & VQ+FT            & NA               & 5K                                                             & 1                                                                         \\ \hline
GGSC \cite{yang2024_GGSC}          & PTC           & GFT              & IGFT             & NA                                                             & $>1$                                                                  \\ \hline
SPZ \cite{spz}           & PTC           &  Quant
+Gzip&         Gzip         & NA                                                             & $>1$                                                                   \\ \hline
MGS \cite{xie2024_mesongs}         & HC            & RAHT             & RAHT+FT          & 8K                                                             & 1                                                                         \\ \hline
ours             & HC            & GPCC+FT          & GPCC             & 10K                                                            & $>1$                                                                   \\ \hline
\end{tabular}
\end{adjustbox}
\end{table}

\section{Conclusion}
\label{sec:conclusion}
We introduced a hybrid compression framework for 3DGS data based on the GPCC codec. This framework integrates adaptive voxelization and attribute refinement to improve compression efficiency while preserving rendering quality. Adaptive voxelization dynamically adjusts resolution based on Gaussian volume and spatial redundancy, reducing the number of Gaussians. Following voxelization, fine-tuning voxelized 3DGS with efficient initialization restores rendering fidelity. Voxelized 3DGS is then efficiently compressed using octree-based lossless encoding for positions, transform coding for spherical harmonics and opacity, and either vector quantization or transform coding for covariance matrices. This scalable compression pipeline enables flexible bitrate adaptation through adjustable quantization parameters, making it well-suited for practical applications such as real-time 3D streaming.

\clearpage
\bibliographystyle{ieeetr}
\bibliography{refs}

\end{document}